\documentclass[a4paper,11pt]{article}
\usepackage{pos}

\usepackage{graphicx} 
\usepackage{amsmath} 
\usepackage{wasysym} 
\usepackage{slashed} 
\usepackage{bigstrut} 
\usepackage{mathtools} 
\usepackage{multirow} 
\usepackage[utf8]{inputenc}
\usepackage{lineno}
\usepackage{todonotes}
\usepackage{xspace}
\usepackage{lscape}
\usepackage[calc,useregional]{datetime2}

\newcommand{\chisq}{\ensuremath{\chi^{2}}\xspace}
\newcommand{\afb}{\ensuremath{A_{\text{FB}}}\xspace}

\newcommand{\neww}[1]{{#1}}

\title{xFitter Updates: Probing $Z$ Boson Couplings with Forward-Backward Asymmetry}
 \ShortTitle{xFitter Updates: Probing $Z$ Boson Couplings with AFB}

\author*[a]{{Oleksandr Zenaiev} (on behalf of the xFitter developers' team)}
\affiliation[a]{II. Institut f\"ur Theoretische Physik, Universit\"at Hamburg \\
	Luruper Chaussee 149, D--22761 Hamburg, Germany}

\emailAdd{oleksandr.zenaiev@desy.de}

\abstract{We present recent updates in the xFitter software framework for global fits of parton distribution functions in high-energy physics. Our focus is on investigating the sensitivity to $Z$ boson couplings using the forward-backward asymmetry in Drell-Yan production. By utilizing an effective approach and simulated data, we assess the accuracy of these couplings, specifically considering the full LHC data sample. Furthermore, we compare our results with predictions for future colliders, providing insights into their potential impact on understanding $Z$ boson interactions.}

\FullConference{31st International Workshop on Deep Inelastic Scattering (DIS2024)\\
 8–12 April 2024\\
Grenoble, France\\}

\begin{document}
\maketitle

The  physics program centering 
on  electroweak precision observables receives 
essential inputs from  measurements of $W$ 
and $Z$ bosons at the LHC.  
Owing  to the cancellation of 
many systematic uncertainties, 
the  forward-backward asymmetry $A_{\rm{FB}}$ 
in Neutron Current Drell-Yan (NCDY) lepton-pair production 
is a crucial component of this program. 
In our paper~\cite{Anataichuk:2023elk} we concentrate on $A_{\rm{FB}}$ asymmetry measurements in NCDY production 
in the region near the $Z$-boson mass scale. The analysis is performed in the framework of the SMEFT 
Lagrangian, including operators up to dimension 
$D = 6$~\cite{Buchmuller:1985jz}, 
\begin{equation} 
	\label{dim6lagr} 
	{\cal L}
	=    {\cal L}^{\rm{(SM)}} + 
	{1 \over \Lambda^2}   \sum_{j=1}^{N_6} C_j^{\rm{(6)}} \ {\cal O}_j^{\rm{(6)}}    \; ,   
\end{equation} 
where the first term on the right hand side is the SM lagrangian, consisting of 
operators of mass dimension $D=4$, while 
the next term is the EFT contribution  containing  
$N_6$ operators $  {\cal O}_j$ of mass dimension $D = 6$, each 
weighted by the dimensionless Wilson coefficient $C_j$ divided by 
$\Lambda^2$, where  $\Lambda$ is  the  ultraviolet mass scale of the EFT. 
In the di-lepton mass region near the  $Z$-boson  
peak, four-fermion operators and 
dipole operators coupling fermions and vector bosons 
can be neglected~\cite{Breso-Pla:2021qoe} in 
Eq.~(\ref{dim6lagr}), and 
the whole  effect of the $D = 6$ SMEFT Lagrangian is a modification of the 
vector boson couplings to fermions, i.e.\ the SMEFT couplings are obtained from the SM couplings via 
the corrections $\delta g$:  
\begin{eqnarray} 
	\label{gLR_SMEFT}
	&&    g_L^{Zu}  \equiv  g_{L ({\rm{SMEFT}})}^{Zu}
	=  g_{L ({\rm{SM}})}^{Zu} + \delta g_L^{Zu}
	\; , \;\;\;\;  g_R^{Zu} \equiv    g_{R ({\rm{SMEFT}})}^{Zu}  =  
	g_{R ({\rm{SM}})}^{Zu}
	+ \delta g_R^{Zu}   \; , 
	\nonumber\\ 
	&&   g_L^{Zd}  \equiv  g_{L ({\rm{SMEFT}})}^{Zd}   = 
	g_{L ({\rm{SM}})}^{Zd}
	+ \delta g_L^{Zd}
	\; , \;\;\;\;    g_R^{Zd}  \equiv  g_{R ({\rm{SMEFT}})}^{Zd} = 
	g_{R ({\rm{SM}})}^{Zd} 
	+ \delta g_R^{Zd} \; . 
\end{eqnarray}
Here $g_{{R(L)} ({\rm{SM}})}^{Zd}$ and $g_{{R(L)} ({\rm{SM}})}^{Zu}$ are the $Z$-boson couplings to right-handed (left-handed) $u$- and $d$-type quarks, respectively.

To perform this study, we use the open-source QCD fit framework \texttt{xFitter}~\cite{Alekhin:2014irh,xFitter:2022zjb} which has been developed for the determination of parton distribution functions (PDFs) and the extraction of fundamental parameters of the SM. It also provides a common framework for the comparison of different theoretical approaches and can be used to test the impact of new
experimental data on the PDFs and on the SM parameters.
We extend the implementation of 
the $A_{\rm{FB}}$~\cite{Accomando:2019vqt} 
in the  \texttt{xFitter} platform
to i) include the SMEFT couplings,
and ii) upgrade the calculations to double-differential 
distributions in both invariant mass $M_{\ell\ell}$ and rapidity $y_{\ell\ell}$ of the 
di-lepton final-state system. 
The largest sensitivity to the couplings is expected in the region $55 \lesssim M_{ll} \lesssim 110$~GeV, and in our analysis we have adopted a slightly wider range $45 < M_{ll} < 145$~GeV. 
\neww{Possible contributions from four-fermion operators in this kinematic range are expected at the $\lesssim 1\permil$ level~\cite{Boughezal:2023nhe}. As a simplifying assumption, we neglected them in our analysis.}

The pseudodata sets are fitted with the four modifications to the couplings $\delta g_L^{Zu}$, $\delta g_R^{Zu}$, $\delta g_L^{Zd}$, $\delta g_R^{Zd}$ being free parameters. We treat the PDF uncertainties using the so-called profiling technique~\cite{Paukkunen:2014zia,HERAFitterdevelopersTeam:2015cre}. In this method, the PDF uncertainties are included in the \chisq using nuisance parameters which are further constrained according to the tolerance criterion of the fit. 

\begin{figure}[!htbp]
	\centering
	\includegraphics[width=0.84\textwidth]{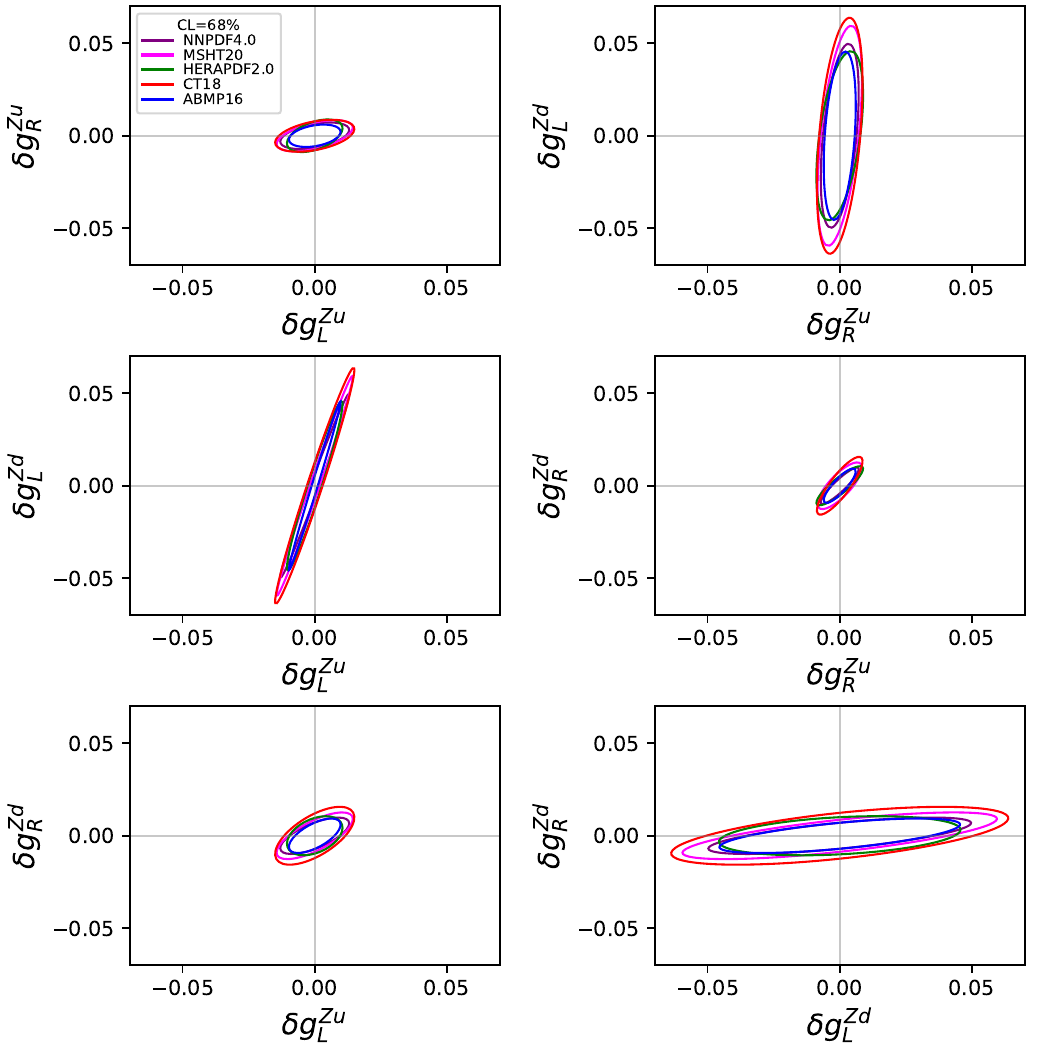}
	\caption{Allowed regions for all pairs of corrections to the $Z$ couplings to $u$- and $d$-type quarks obtained using different PDF sets.}
	\label{fig:ell_pdf}
\end{figure}

In Fig.~\ref{fig:ell_pdf} the allowed regions for the couplings are shown as obtained using different PDF sets~\cite{Hou:2019efy,Bailey:2020ooq,NNPDF:2021njg,Alekhin:2017kpj,H1:2015ubc}. 
Both the size and the shape of the allowed regions are similar, independent of the PDF set.

\begin{figure}[!htbp]
	\centering
	\includegraphics[width=1.00\textwidth]{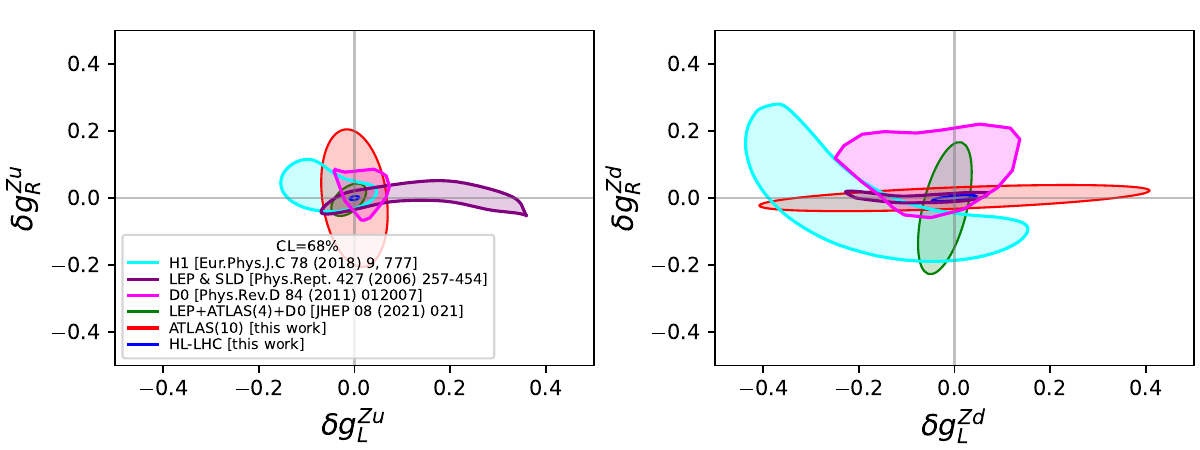}
	\caption{Allowed regions for all pairs of corrections to the $Z$ couplings to $u$- and $d$-type quarks obtained using HL-LHC pseudodata as well as different existing data sets.}
	\label{fig:ell_past}
\end{figure}

In Fig.~\ref{fig:ell_past} we compare our results obtained for the HL-LHC\footnote{We show our results obtained using the ABMP16 PDF set, since only this set provides symmetric PDF errors which are easier to include in the PDF profiling.} with the other analyses of existing data from LEP, Tevatron, HERA and LHC. Namely, we compare with the analysis of the H1 Collaboration at HERA~\cite{H1:2018mkk}, the LEP+SLD combination~\cite{ALEPH:2005ab}, the analysis of D0 Collaboration at Tevatron~\cite{D0:2011baz} and the analysis of LEP, ATLAS and D0 data from Ref.~\cite{Breso-Pla:2021qoe}. %
In addition to the HL-LHC results, we present our results of analyzing all available 10 bins from the ATLAS measurement of \afb~\cite{ATLAS:2018gqq}, while only $4$ bins at the $Z$ peak were used in the analysis of Ref.~\cite{Breso-Pla:2021qoe}. 
The level of precision expected at the HL-LHC outperforms any existing data sets.

\begin{figure}[!htbp]
	\centering
	\includegraphics[width=0.84\textwidth]{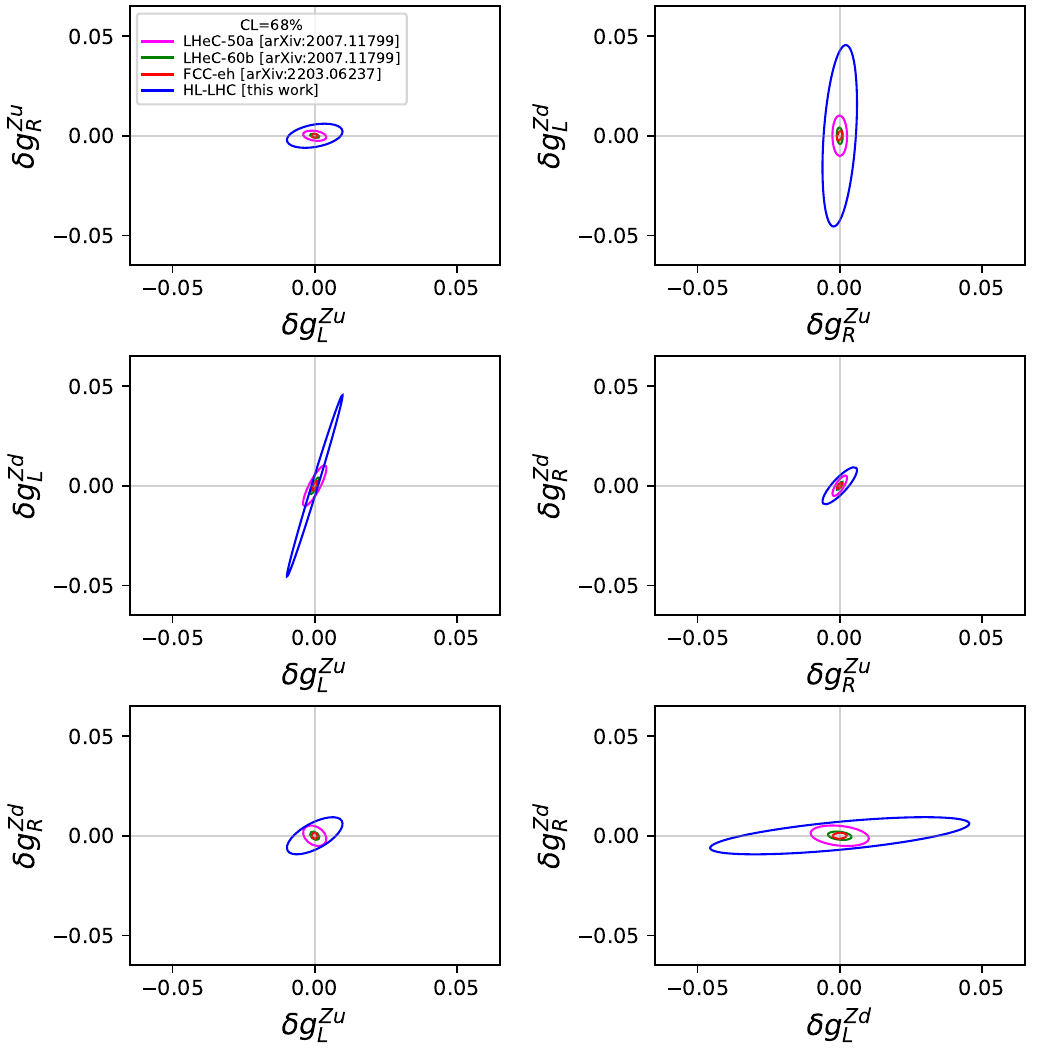}
	\caption{Allowed regions for all pairs of corrections to the $Z$ couplings to $u$- and $d$-type quarks obtained using HL-LHC pseudodata compared to the ones for different future experiments.}
	\label{fig:ell_future}
\end{figure}

\begin{figure}[!htbp]
	\centering
	\includegraphics[width=0.495\textwidth]{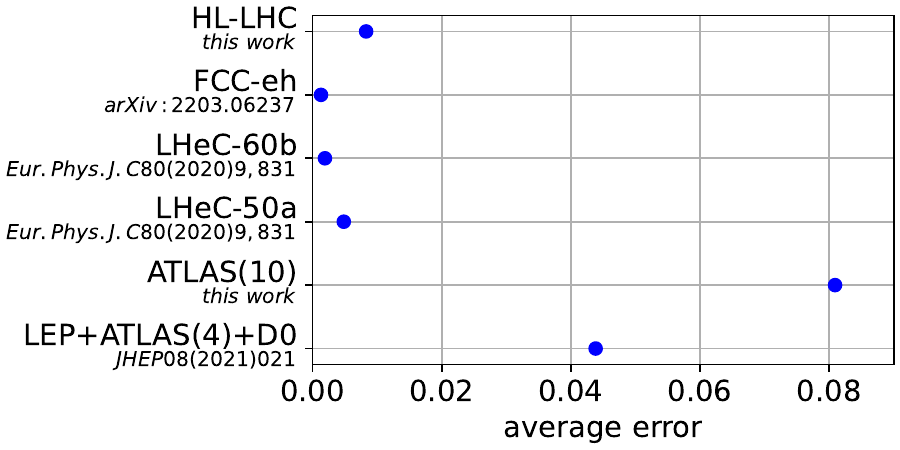}
	\includegraphics[width=0.495\textwidth]{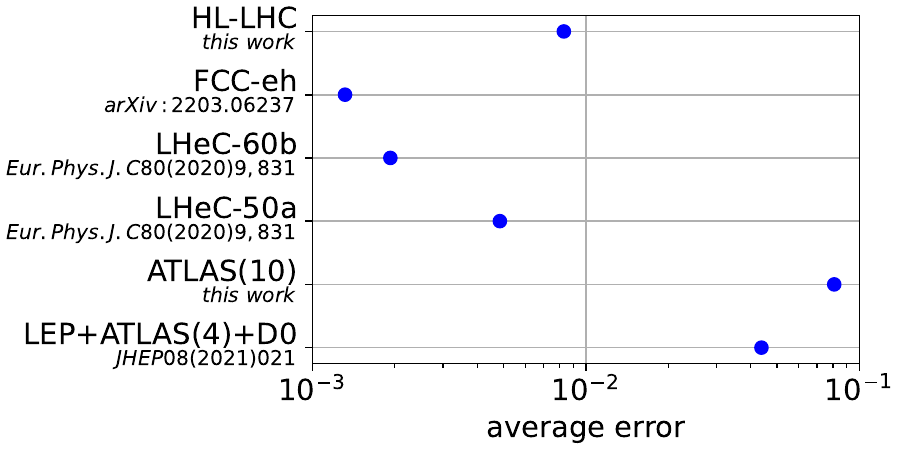}
	\caption{The average size of the uncertainties on the fitted corrections to the $Z$ couplings to $u$- and $d$-type quarks for different future experiments using the linear (left) or logarithmic (right) scale.}
	\label{fig:aver}
\end{figure}

In Fig.~\ref{fig:ell_future} we compare the results obtained for the HL-LHC with the results expected at the future colliders currently under discussion,  LHeC~\cite{Britzger:2020kgg} and FCC-eh~\cite{Britzger:2022abi}. 
Furthermore, in Fig.~\ref{fig:aver} the average size of the uncertainties which can be obtained using current and future data sets are compared. A sub-percent level of precision is expected at the LHeC, FCC-eh and HL-LHC, which is one order of magnitude better than what can be obtained using existing data sets from LEP, Tevatron, HERA and LHC. 

In summary, we have studied the possibility to improve constraints on the $Z$ couplings to the $u$- and $d$-type quarks using the future measurements of \afb at the HL-LHC. 
The results were compared with the existing analyses of the LEP, HERA, Tevatron and LHC data, as well as with the results which are expected at the future colliders LHeC and FCC-eh. The uncertainties on the $Z$ couplings to the $u$- and $d$-type quarks at the HL-LHC are expected at percent level, thus outperforming by an order of magnitude any determinations of these couplings using existing data sets. This level of precision is similar, but a little inferior to the one which is expected at the LHeC and FCC-eh.

\bibliographystyle{JHEP} 
\bibliography{afbsmeft}

\providecommand{\href}[2]{#2}\begingroup\raggedright\begin{thebibliography}{10}

\bibitem{Anataichuk:2023elk}
A.~Anataichuk et~al., \emph{{Exploring SMEFT Couplings Using the
  Forward-Backward Asymmetry in Neutral Current Drell-Yan Production at the
  LHC}},  \href{https://arxiv.org/abs/2310.19638}{{\ttfamily 2310.19638}}.

\bibitem{Buchmuller:1985jz}
W.~Buchmuller and D.~Wyler, \emph{{Effective Lagrangian Analysis of New
  Interactions and Flavor Conservation}},
  \href{https://doi.org/10.1016/0550-3213(86)90262-2}{\emph{Nucl. Phys. B}
  {\bfseries 268} (1986) 621}.

\bibitem{Breso-Pla:2021qoe}
V.~Bres\'o-Pla, A.~Falkowski and M.~Gonz\'alez-Alonso, \emph{{A$_{FB}$ in the
  SMEFT: precision Z physics at the LHC}},
  \href{https://doi.org/10.1007/JHEP08(2021)021}{\emph{JHEP} {\bfseries 08}
  (2021) 021} [\href{https://arxiv.org/abs/2103.12074}{{\ttfamily
  2103.12074}}].

\bibitem{Alekhin:2014irh}
S.~Alekhin et~al., \emph{{HERAFitter}},
  \href{https://doi.org/10.1140/epjc/s10052-015-3480-z}{\emph{Eur. Phys. J. C}
  {\bfseries 75} (2015) 304} [\href{https://arxiv.org/abs/1410.4412}{{\ttfamily
  1410.4412}}].

\bibitem{xFitter:2022zjb}
{\scshape xFitter} collaboration, \emph{{xFitter: An Open Source QCD Analysis
  Framework. A resource and reference document for the Snowmass study}},  6,
  2022 [\href{https://arxiv.org/abs/2206.12465}{{\ttfamily 2206.12465}}].

\bibitem{Accomando:2019vqt}
E.~Accomando et~al., \emph{{PDF Profiling Using the Forward-Backward Asymmetry
  in Neutral Current Drell-Yan Production}},
  \href{https://doi.org/10.1007/JHEP10(2019)176}{\emph{JHEP} {\bfseries 10}
  (2019) 176} [\href{https://arxiv.org/abs/1907.07727}{{\ttfamily
  1907.07727}}].

\bibitem{Boughezal:2023nhe}
R.~Boughezal, Y.~Huang and F.~Petriello, \emph{{Impact of high invariant-mass
  Drell-Yan forward-backward asymmetry measurements on SMEFT fits}},
  \href{https://doi.org/10.1103/PhysRevD.108.076008}{\emph{Phys. Rev. D}
  {\bfseries 108} (2023) 076008}
  [\href{https://arxiv.org/abs/2303.08257}{{\ttfamily 2303.08257}}].

\bibitem{Paukkunen:2014zia}
H.~Paukkunen and P.~Zurita, \emph{{PDF reweighting in the Hessian matrix
  approach}}, \href{https://doi.org/10.1007/JHEP12(2014)100}{\emph{JHEP}
  {\bfseries 12} (2014) 100} [\href{https://arxiv.org/abs/1402.6623}{{\ttfamily
  1402.6623}}].

\bibitem{HERAFitterdevelopersTeam:2015cre}
{\scshape HERAFitter developers' Team} collaboration, \emph{{QCD analysis of
  $W$- and $Z$-boson production at Tevatron}},
  \href{https://doi.org/10.1140/epjc/s10052-015-3655-7}{\emph{Eur. Phys. J. C}
  {\bfseries 75} (2015) 458}
  [\href{https://arxiv.org/abs/1503.05221}{{\ttfamily 1503.05221}}].

\bibitem{Hou:2019efy}
T.-J.~Hou et~al., \emph{{New CTEQ global analysis of quantum chromodynamics
  with high-precision data from the LHC}},
  \href{https://doi.org/10.1103/PhysRevD.103.014013}{\emph{Phys. Rev. D}
  {\bfseries 103} (2021) 014013}
  [\href{https://arxiv.org/abs/1912.10053}{{\ttfamily 1912.10053}}].

\bibitem{Bailey:2020ooq}
S.~Bailey, T.~Cridge, L.A.~Harland-Lang, A.D.~Martin and R.S.~Thorne,
  \emph{{Parton distributions from LHC, HERA, Tevatron and fixed target data:
  MSHT20 PDFs}},
  \href{https://doi.org/10.1140/epjc/s10052-021-09057-0}{\emph{Eur. Phys. J. C}
  {\bfseries 81} (2021) 341}
  [\href{https://arxiv.org/abs/2012.04684}{{\ttfamily 2012.04684}}].

\bibitem{NNPDF:2021njg}
{\scshape NNPDF} collaboration, \emph{{The path to proton structure at 1\%
  accuracy}}, \href{https://doi.org/10.1140/epjc/s10052-022-10328-7}{\emph{Eur.
  Phys. J. C} {\bfseries 82} (2022) 428}
  [\href{https://arxiv.org/abs/2109.02653}{{\ttfamily 2109.02653}}].

\bibitem{Alekhin:2017kpj}
S.~Alekhin, J.~Bl\"umlein, S.~Moch and R.~Placakyte, \emph{{Parton distribution
  functions, $\alpha_s$, and heavy-quark masses for LHC Run II}},
  \href{https://doi.org/10.1103/PhysRevD.96.014011}{\emph{Phys. Rev. D}
  {\bfseries 96} (2017) 014011}
  [\href{https://arxiv.org/abs/1701.05838}{{\ttfamily 1701.05838}}].

\bibitem{H1:2015ubc}
{\scshape H1, ZEUS} collaboration, \emph{{Combination of measurements of
  inclusive deep inelastic ${e^{\pm }p}$ scattering cross sections and QCD
  analysis of HERA data}},
  \href{https://doi.org/10.1140/epjc/s10052-015-3710-4}{\emph{Eur. Phys. J. C}
  {\bfseries 75} (2015) 580}
  [\href{https://arxiv.org/abs/1506.06042}{{\ttfamily 1506.06042}}].

\bibitem{H1:2018mkk}
{\scshape H1} collaboration, \emph{{Determination of electroweak parameters in
  polarised deep-inelastic scattering at HERA}},
  \href{https://doi.org/10.1140/epjc/s10052-018-6236-8}{\emph{Eur. Phys. J. C}
  {\bfseries 78} (2018) 777}
  [\href{https://arxiv.org/abs/1806.01176}{{\ttfamily 1806.01176}}].

\bibitem{ALEPH:2005ab}
{\scshape ALEPH, DELPHI, L3, OPAL, SLD, LEP Electroweak Working Group, SLD
  Electroweak Group, SLD Heavy Flavour Group} collaboration, \emph{{Precision
  electroweak measurements on the $Z$ resonance}},
  \href{https://doi.org/10.1016/j.physrep.2005.12.006}{\emph{Phys. Rept.}
  {\bfseries 427} (2006) 257}
  [\href{https://arxiv.org/abs/hep-ex/0509008}{{\ttfamily hep-ex/0509008}}].

\bibitem{D0:2011baz}
{\scshape D0} collaboration, \emph{{Measurement of $\sin^2\theta_{\rm
  eff}^{\ell}$ and $Z$-light quark couplings using the forward-backward charge
  asymmetry in $p\bar{p} \to Z/\gamma^{*} \to e^{+}e^{-}$ events with ${\cal
  L}=5.0$ fb$^{-1}$ at $\sqrt{s}=1.96$ TeV}},
  \href{https://doi.org/10.1103/PhysRevD.84.012007}{\emph{Phys. Rev. D}
  {\bfseries 84} (2011) 012007}
  [\href{https://arxiv.org/abs/1104.4590}{{\ttfamily 1104.4590}}].

\bibitem{ATLAS:2018gqq}
{\scshape ATLAS} collaboration, \emph{{Measurement of the effective leptonic
  weak mixing angle using electron and muon pairs from $Z$-boson decay in the
  ATLAS experiment at $\sqrt s = 8$ TeV}}, {\emph{ATLAS-CONF-2018-037} (2018)
  }.

\bibitem{Britzger:2020kgg}
D.~Britzger, M.~Klein and H.~Spiesberger, \emph{{Electroweak physics in
  inclusive deep inelastic scattering at the LHeC}},
  \href{https://doi.org/10.1140/epjc/s10052-020-8367-y}{\emph{Eur. Phys. J. C}
  {\bfseries 80} (2020) 831}
  [\href{https://arxiv.org/abs/2007.11799}{{\ttfamily 2007.11799}}].

\bibitem{Britzger:2022abi}
D.~Britzger, M.~Klein and H.~Spiesberger, \emph{{Precision electroweak
  measurements at the LHeC and the FCC-eh}},
  \href{https://doi.org/10.22323/1.398.0485}{\emph{PoS} {\bfseries EPS-HEP2021}
  (2022) 485} [\href{https://arxiv.org/abs/2203.06237}{{\ttfamily
  2203.06237}}].

\end{thebibliography}\endgroup

\end{document}